\documentclass[rmp,twocolumn,a4paper,showkeys,superscriptaddress]{revtex4}
\usepackage[latin1]{inputenc}
\usepackage{graphicx}

\begin{document}

\title{Adaptive walks in a gene network model of morphogenesis: \\
insights into the Cambrian explosion}

\providecommand{\ICREA}{ICREA-Complex Systems Lab, Universitat Pompeu Fabra (GRIB), 
Dr Aiguader 80, 08003 Barcelona, Spain}
\providecommand{\SFI}{Santa Fe Institute, 1399 Hyde Park Road, Santa Fe NM 87501, USA}
\providecommand{\UNM}{Department of Cell Biology and Physiology, Health Science Center, 
University of New Mexico, \\ 
Albuquerque, NM 87131 USA}

\author{Ricard V. Solé}
\affiliation{\ICREA}
\affiliation{\SFI}
\author{Pau Fernández}
\affiliation{\ICREA}
\author{Stuart A. Kauffman}
\affiliation{\SFI}
\affiliation{\UNM}

\begin{abstract}
  The emergence of complex patterns of organization close to the
  Cambrian boundary is known to have happened over a (geologically)
  short period of time. It involved the rapid diversification of body
  plans and stands as one of the major transitions in evolution.  How
  it took place is a controversial issue. Here we explore this problem
  by considering a simple model of pattern formation in multicellular
  organisms. By modeling gene network-based morphogenesis and its
  evolution through adaptive walks, we explore the question of how
  combinatorial explosions might have been actually involved in the
  Cambrian event. Here we show that a small amount of genetic
  complexity including both gene regulation and cell-cell signaling
  allows one to generate an extraordinary repertoire of stable spatial
  patterns of gene expression compatible with observed anteroposterior
  patterns in early development of metazoans. The consequences for the  understanding of the 
tempo and mode of the Cambrian event are outlined. 
\end{abstract}


\maketitle

\section{Introduction}

Morphological complexity in metazoa experienced an extraordinary leap
close to the Cambrian boundary  (around 550 Myrs ago). Such event has been labeled the {\em
  Cambrian explosion}. When this radiation began, and how rapidly it
unfolded, is still the subject of active research \citep{Morris,Raff,Valentine99,Solecambrian}. 
The early origins of the last common ancestor and the structure of its bodyplan are controversial 
issues \citep{Erwin2002}.
The Cambrian event established essentially all the major
animal body plans and hence all the major phyla which would exist
thereafter. A body plan can be described anatomically but also in
terms of the spatiotemporal pattern of expression of some key genes
\citep{Arthur}. In this context, the origins of evolutionary
novelty emerges as an integrative field involving gene regulation,
development and paleobiology \citep{Arthur2,Carroll2001}. As Shubin and Marshall pointed
out, untangling the web of ecological, developmental and genetic
interactions is a difficult task and a key question is which changes
occur first \citep{Shubin}. 

A possible interpretation of the uniqueness, tempo and mode of the Cambrian event in terms of generic
features of complex evolving systems was suggested by Kauffman in
terms of adaptive walks on rugged fitness landscapes
\citep{Kauffman1993,Kauffman1989}. In a nutshell, 
the idea is that, given the fundamental constraints imposed by early 
developmental dynamics, early exploration of the universe of possible 
body plans took place quickly (once some underlying requirements were met) 
but then slowed down as the repertoire of possible bodyplans was filled out \citep{Kauffman1989,Raff}. 
Starting from some initial condition where low-fit multicellular
organisms were present, a rapid exploration of the
landscape was allowed to occur.  This initial exploration led to an
increase of diversity of improved alternative morphologies thereby
establishing phyla. As the rate of finding fitter mutants that alter
early developmental processes (which define body plans) slowed down, lower taxonomic
groups became established. 

\begin{figure*}
  {\centering \includegraphics[width=12cm]{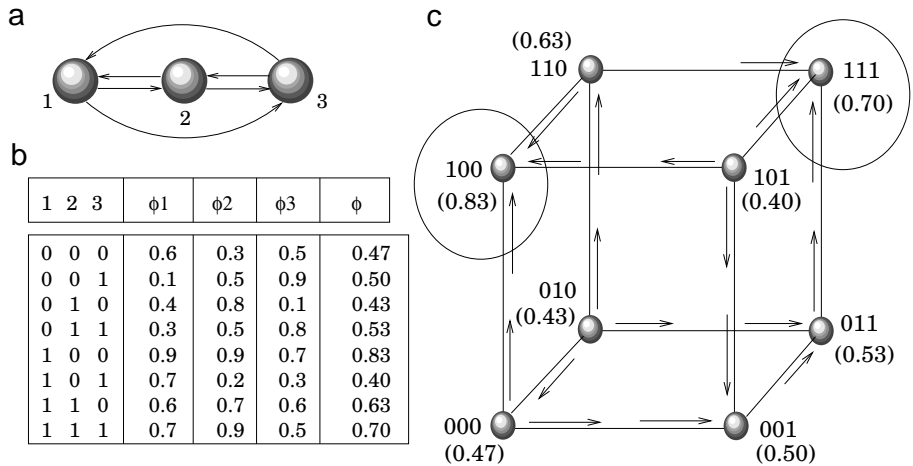}}
\caption{
  \label{cubefig}
  A simple fitness landscape with $N=3$ traits and $K=2$ epistatic interactions. Here the three 
traits involved interact with the other two (a). The possible states of the system are 
  given by the vertices of the fitness cube (this time a
  $N=3$-dimensional cube $\Gamma_3$). The resulting average fitness, $\phi=\sum \phi_i/N$ 
is obtained from a fitness table (b) where the contribution of each 
trait $\phi_i$ (under the presence of the other two) is generated as a random 
number between zero and one. The table provides the local maxima.
  Adaptive walks (indicated as arrows) take place to nearest sites with higher fitness. In
  this example, adaptive walks end in two possible local peaks (c).}
\end{figure*}

The argument is completed by the assumption that mutants affecting early development have
more profound effects than those affecting late development. In this
scenario, by the Permian extinction (200 Myr later), when 
an estimated $95 \%$ of species ($82 \%$ of genera, see \citep{ErwinPermian}) went extinct, 
early developmental pathways would
be expected to have become largely frozen in after the Cambrian event and no new phyla 
would be reachable. Fitter variants altering
basic body plans would be very hard to find but not variants affecting
late development. This would allow the radiation of new families and lower taxa.
Fitness landscapes, first introduced by Wright in 1932, allow to describe
changes in the fitness of a given species through time in a
well-defined fashion \citep{Palmer,Kauffman1993,Rowe,StadlerLands}. In an idealized situation, we can imagine the
landscape as a surface.  Here the tops and bottoms of the landscape
would indentify good and bad combinations of traits, respectively. This
picture allows one to visualize the evolution of species as a hill climbing on the fitness surface.

A more precise definition is provided by
describing each individual (or species) is defined as a set of $N$ binary variables $S_i\in\{0,1\}$ 
defining a set of characteristic traits \citep{KauffmanLevin,Kauffman1993}. The total number of
combinations is $2^N$ and each occupies a node of a $N$-hypercube
$\Gamma_N$. In figure \ref{cubefig} we show an example of this fitness
landscape for $N=3$.  Each string has a well-defined fitness value, which
can be represented by means of a $N$-dimensional function $\phi =
\phi(S_1, ..., S_N)$. If the number of epistatic interactions $K$ is zero, 
i. e. if different traits do not influence each other, then a smooth landscape 
is obtained, whith a single peak. But if different traits interact, then the landscape 
becomes rugged and multiple local fitness peaks are allowed to occur.  
The simplest evolution on a fitness cube occurs by
means single, one-bit steps. In other words, a given species can perform a random {\em adaptive walk} from a given node towards one of
its $N$ nearest neighbors if this leads to an increase in fitness (alternatively, 
a neutral change can also occur, and thus random drift is also allowed). 
A direct consequence of this assumption is that once a local peak is reached, no further changes are allowed. 


There is a universal feature of adaptation on statistically
correlated landscapes \citep{KauffmanLevin} which can be 
appropriately used to test these ideas.  This can be formulated as
follows: if $S$ is the cumulative number of assumed
improvements (new body plans) originated and $T$ is the cumulative
number of tries, then $S$ will grow with
$T$ in a logarithmic fashion. Graphically, this means that $S$ grows
rapidly at the beginning and then slows down. The cumulative number of
tries can be approximated by the cumulative number of lower-level
entities (e. g. genera) viewed as successive experiments in the
generation of higher-level entities. The analysis of the cumulative increase of 
phylum originations against cumulative genera seems to confirm this prediction \citep{Eble}.

A further step in exploring the possible scenarios for an explosion 
of patterns in the evolution of development should consider the 
developmental program in an explicit way. 
One important factor not addressed by previous theoretical studies
deals with the spectrum of possible spatial patterns of gene
expression that can be generated from a given number of genes. This is
a relevant question since combinatorial explosions can occur once
complexity thresholds (in number of genes or their interactions)
are reached.  In previous studies, it has been shown that some
particular types of combinations of gene-gene interactions involving
cell-cell communication allow to easily generate a number of spatial
patterns including stripes, gradients or spots \citep{SalazarJTB,SalazarEVOLDEV,SoleFYSA}. These studies were
performed on randomly wired networks and revealed a large fraction of
spatial patterns that could be generated and their nature \citep{SoleFYSA}. One
question immediately emerges: what are the minimal complexity
requirements in terms of number of pattern-forming genes- in order to
reach a high diversity of phenotypes?

One possible approach to the problem is to consider an artificial
evolution model were real organisms and their development are replaced
by digital organisms \citep{WilkeTREE} with simplified developmental programs. Such an
approximation is becoming more common in evolutionary studies. 
The success of some of these models in extraordinary in providing 
insight into the evolution of complex biological systems. As
pointed out by Gould, such a range of success is a consequence of
universal laws of change that are common to all complex systems
\citep{GouldBible}. In this context, we can conjecture that digital
developmental programs might also share generic properties with real,
early development.


Previous theoretical studies have shown that the
fact of randomly wiring gene networks within a tissue context allows a
high diversity of patterns and that spatial patterns are easily
obtained \citep{SoleFYSA}. But no theoretical study has addressed the question of
the potential diversity of (stable) patterns that can be obtained by
tuning the number of genes involved in creating them, provided that
some simple initial signal (an activated gene) is present at the
beginning of development. Here we will explore this question by means
of an simplified model of early development in which the number of
different cell types is optimized.  In this context, we specifically
ask the following questions:

\begin{enumerate}
  
\item What are the consequences of searching for stable patterns of
  increasing complexity, being complexity measured in terms of the
  number of cell types?
  
\item What are the minimal requirements in terms of gene regulation
  complexity in order to be able to reach a wide spectrum of spatial
  patterns?
  
\item What type of spatial patterns are obtained?
  
\item How does the evolution of these patterns takes place in terms of
  the underlying fitness landscape?
  
\item Can a combinatorial explosion partially explain the tempo and
  mode of the Cambrian event?

\end{enumerate} 

As will be shown below, the model approach provides nontrivial, tentative 
answers to the previous questions.

\section{Gene network model}

\begin{figure}
\begin{center}
  {\centering\includegraphics[width=8cm]{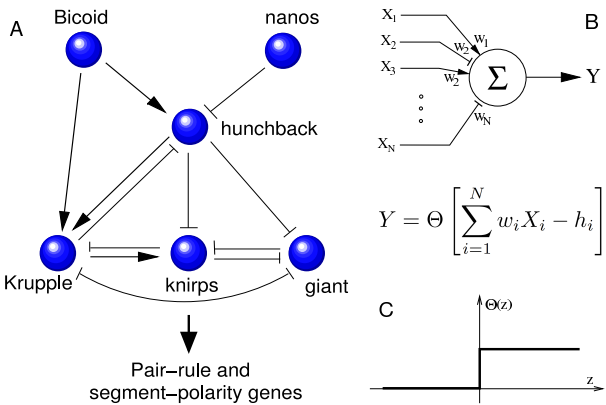}}
\end{center}
\caption{
  \label{earlygenesfig}
  (A) Gene interactions in early development. These is a subset of gene regulatory 
  interactions that take place in {\em Drosophila} development. Arrows
  indicate different types of positive and negative interactions among
  different genes. In (B) the simplified, discrete threshold model
  considered here is shown. Each gene is treated as a binary (Boolean)
  variable with only two states: active (1) or inactive (0). It
  integrates the input signals which are weighted through a matrix of
  links $W$ (either positive or negative). The output $Y$ is obtained
  by using a threshold function, such as the one indicated in (C).}
\end{figure}

A complete model of the gene activity pattern even at early stages of
development would require a detailed description of the
different levels of gene regulation and signaling \citep{Davidson}. Such a description
should take into account the continuous nature of mRNA and
protein levels as well as other considerations related to the nature
and distribution of cell signaling molecules and the stochastic nature
of their dynamics. In this context, abstract models only taking into account a
small amount of key features often capture the essential ingredients
of gene regulation dynamics \citep{Hasty,SoleSatorras2002}. 

As a matter of fact, it has been recently shown that discrete, ON-OFF
models of early development can actually provide complete enough
information in order to reproduce the key traits of a developmental
pattern \citep{Bodnar,Mendoza,Sanchez,Albert2003}. 
As discussed in \citep{Albert2003} within the context of the
expression pattern of segment polarity genes in {\em Drosophila}, the
observed patterns are determined by the topology of the network and
the type of regulatory interactions between components. This is consistent 
with recent computer models indicating that a robust segment polarity module 
exists and that it is rather insensitive to variation in kinetic parameters \citep{vonDassow2000} 
(see also \citep{Meir,Gibson}).
In other words, a complete description of both wild type patterns and
various mutants is successfully reached by using simple Boolean
networks. Here we take the same approach.

\begin{figure}
\begin{center}
  {\centering \includegraphics[width=7.5cm]{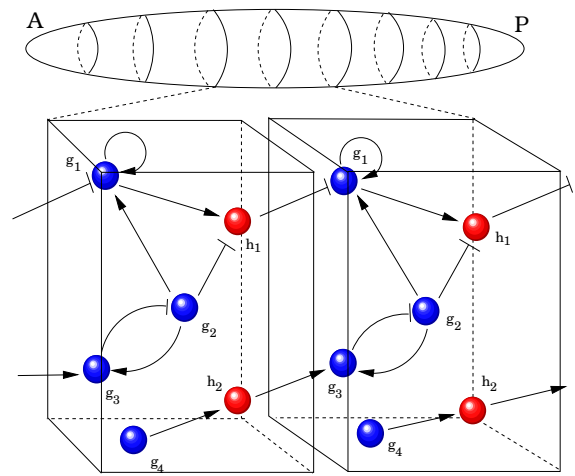}}
\end{center}
\caption{
  \label{boolcellsfig}
  Modeling morphogenesis using simple gene network models. Here 
  the simulated organism is a one-dimensional array of cells. Each cell
  contains the same set of $N$ genes, $G$ of which (indicated by blue
  balls) only interact inside the cell whereas $H$ others (red balls)
  act as microhormones, exchanging information with neighboring
  cells.}
\end{figure}

Several choices can be made in dealing with the structure of the 
wiring scheme to be used in the regulation network. One possibility 
is to restrict ourselves to a hierarchical cascade of interactions inspired 
in the topology of interactions of {\em Drosophila}. 
But it is known that even gap genes are not completely hierarchical \citep{Gehring}. 
Some of them may act to switch on others  (figure 2a). Actually, this
property characterizes other gene groups as well, reminding us that we
are actually dealing with a complex network instead of a linear chain
of steps \citep{Arthur}. 

The model explored here is a gene threshold system, described by a set
of $N$ genes per cell, interacting through a one-dimensional domain
involving $C$ cells, as shown in figure \ref{boolcellsfig}. The set of genes involved 
represent those required to build the basic structure and are thus bauplan genes (as defined 
in \citep{Tautz}). 
Here gene states will be Boolean: genes are
either active (1) or inactive (0).  Two types of elements will be
considered: $G$ genes and $H$ hormones, with $N=G+H$. Genes interact
within the cell, whose state at a given time $t$ will be indicated as
$g_i^j(t)$, with $i=1, ..., G$ as the gene number and $j=1, ..., C$ as
the cell number (ordered from anterior to posterior). Generically
labeled {\em microhormones} \citep{Jackson} involve some implicit local mediators
communicating neighboring cells.  These hormones can receive inputs from any
of the first $G$ units, but they can only make output to genes in
other cells. The state of these $H$ hormones will be accordingly 
indicated as $h_i^j(t)$.

Two matrices will be required in order to define the whole spectrum of
links between different elements. These two matrices will be indicated
by ${\bf A}=(A_{ik})$ and ${\bf B}=(B_{ik})$, defining interactions
among the $G$ genes and between genes and hormones, respectively.

The basic set of equations of our gene network model are:
\begin{eqnarray}
g_i^j(t+1) &=& \Theta \left [ {\cal{G}}_i^j(t) + {\cal{H}}_i^j(t) \right ] \\
h_i^j(t+1) &=& \Theta \left [ {\cal{G}}_i^j(t) \right ] 
\end{eqnarray}
with
\begin{eqnarray*}
{\cal{G}}_i^j(t) &=& \sum_{k=1}^G A_{ik} g_i^j(t) \nonumber \\
{\cal{H}}_i^j(t) &=& \sum_{k=1}^G B_{ik} \delta \left( h_k^{j+1}(t) , h_k^{j-1}(t) \right) \nonumber
\end{eqnarray*}
where $\delta(x,y)$ is an ``OR'' function (that is, $\delta(x,y)=1$ if either $x=1$ or $y=1$ 
and zero if $x=y=0$). The function $\Theta(x)$ is a threshold function (that is, $\Theta(x)=1$ if
$x>0$ and $\Theta(x)=0$ otherwise), and is shown in figure \ref{earlygenesfig}C. 

Therefore, genes are influenced either by genes or hormones and 
hormones are influenced just by genes, and the influences ${\cal{G}}_i^j$ 
and ${\cal{H}}_i^j$ add up to determine if genes will be active or inactive
by the comparison to a threshold (in this case 0), as shown in figure \ref{earlygenesfig}B. 
The OR function, or $\delta(x,y)$, is used here as a substitute for the
diffusion of microhormone to the neighbours, since a microhormone will 
be seen as active in cell $j$ if either of its neighbours is active. 

For cells at the boundaries we have the special terms:
\begin{eqnarray}
{\cal{H}}_i^1(t) &=& \sum_{k=1}^G B_{ik} h_k^{2}(t) \nonumber \\
{\cal{H}}_i^C(t) &=& \sum_{k=1}^G B_{ik} h_k^{C-1}(t) \nonumber
\end{eqnarray}
indicating that cells at the poles just have one neighbour.

Finally, we need to define an initial condition. The simplest choice
that can be defined involves the activation of a single gene at the
anterior ($j=1$) pole, as well as the symmetric case in which the
signal is present at the two poles ($j=1$ and $j=C$). Specifically, we set
$g_{i}^j(0)=h_{i}^j(0)=0$ for all $j=1, ..., N_C$, except $g_1^1(0)=1$
in the single pole case, and an additional $g_1^{N_C}(0)=1$ in the
symmetric case.
These choices correspond to maternal signals confined to the embryo's
extremes.  Such initial change will propagate to the rest of the
tissue provided that the network is sensitive to it.  

\begin{figure}
\begin{center}
  {\centering \includegraphics[height=13cm]{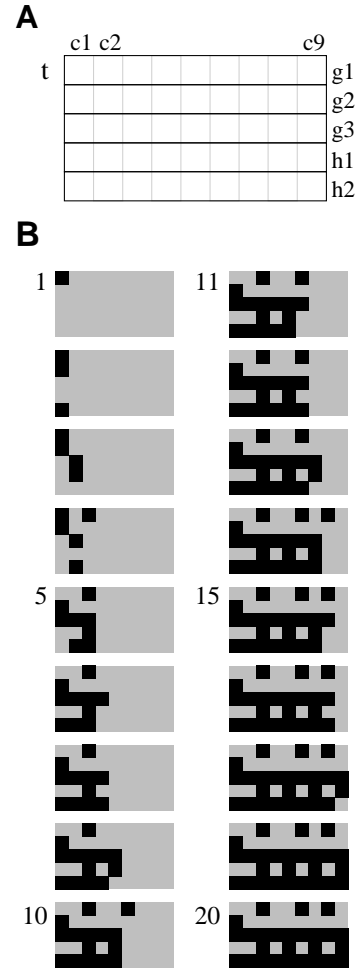}}
\end{center}
\caption{
  \label{temporal}
  The temporal construction of a pattern in a $G=3$ genes and $H=2$ hormones network.
  (A) Scheme of representation. (B) The 20 steps necessary to reach a stable pattern.
  Genes in black are active and genes in gray are inactive.
}
\end{figure}

As an example of the dynamics, the temporal evolution of one sample
organism with one-pole maternal signal is shown in figure
\ref{temporal}.


\section{Adaptive walks}

Using the above mentioned gene network model of pattern formation, we now
explore how pattern complexity emerges through a simple evolutionary
algorithm where a population of $P$ organisms evolve through adaptive
walks. Such type of algorithm has been successfully used in different
contexts (see for example \citep{Niklas1994,Niklas1997}). 



In order to define the evolution algorithm, we need to first define a
fitness function. Here we restrict ourselves to measuring
the number of different cell types. The number of cell types is a good
measure of complexity which is known to increase
through metazoan evolution \citep{Valentine,Carroll2001}.  Increases in cell type number
provide a high potential for further evolution of anatomical and
functional complexity, essentially through division of labor and the
formation of specialized tissues \citep{Maynard}. It is clear that the diversity of organisms
involving a small number of cell types is fairly limited and thus a
first step towards the evolution of complex organisms requires an
expansion of their diversity of cell types. Actually, as discussed by Erwin and 
Davidson, regulatory processes underlying cell-type specification are very old and 
display conserved plesiomorphic features \cite{Erwin2002}. Morphogenetic 
processes would have evolved independently and added at later stages. 

We consider, then, the number of cell types $n_{ct}$ as our
complexity measure and formulate the following question. What are the
consequences for the diversity of spatial patterns that can be
generated if we increase the number of cell types?

Starting from a homogeneous population of $P$ identical organisms, we perform
adaptive walks on the fitness landscape. At each generation in the algorithm, we
sequentially choose each organism and introduce a single change in its
gene network. Three types of changes can occur, with probability
$\frac{1}{3}$, all of them affecting the dependencies between genes:
\begin{itemize}
\item addition of new links,
\item removal of previously present links, and
\item randomization of links' weights.
\end{itemize}
In this way the complexity of the network can be tuned independently
for each organism.

After the change, the developmental pattern generated by the organism
is simulated.  The number of different cell types $n^{t}_{ct}$ is
computed using this new pattern and it is compared with the previous
one, $n^{t-1}_{ct}$. If $n^{t}_{ct} \ge n^{t-1}_{ct}$ the change is
accepted and the new organism replaces the old one. Otherwise the
change is rejected and the old organism is kept.  The fact that we
keep the new organism if $n^{t}_{ct} = n^{t-1}_{ct}$ (the exact
developmental pattern might be different), introduces a certain amount
of neutrality that may reduce the probability that an organism gets
trapped because it cannot find any changes that give it a higher
number of cell types. As an additional requirement, the stability of the final pattern is
enforced by rejecting those organisms which do not reach a stable
state in $I$ iterations.  This is a strong constrain,
since puts a hard limit on the attainable patterns, but it also seems
sensible, for developmental processes are very robust and only display 
transient oscillations.

The size of the space of possible patterns $\Gamma$ is rather high,
since the number of combinations of the activation of genes in each cell
gives rise to a potential number of patterns of
\begin{equation}
\vert \Gamma \vert = 2^{(H + G)C}. 
\end{equation}
Actually, the tissue context exponentially expands the allowed space
$\Gamma_c$ of cell types reachable by isolated cells, which has a size
$\vert \Gamma_c \vert = 2^{(H + G)}$ (i.e. we have $\vert \Gamma
\vert=\vert \Gamma_c \vert^C$).  Even for a small number of cells
and a small amount of genes, the resulting space
is hyperastronomically large. For example, if we take $G + H=4$ (which
is one particularly interesting case, as shown below) we have $\vert
\Gamma \vert = 2^{60} \approx 10^{17}$.

In order to explore the potential repertoire of structures that can be
generated, we will focus in the set of stable spatial patterns
$\Gamma_g$ associated to each single gene (independently of others),
from the set of possible spatial patterns, with $\vert \Gamma_g \vert =
2^C$.  This is equivalent to consider the raw capacity to generate
specific patterns without regarding any gene as more important than
any other.  Therefore, a population of organisms at any given
generation has a corresponding set of patterns, ${\cal{P}}$, the ones
taken from all the patterns that the genes in each organism generate
individually in the developmental process. Any pattern in this list
occurs in some specific genes in some organisms, across the different cells, 
and with a different frequency, which we also measure.

\section{Results}


\begin{figure}
\begin{center}
  \includegraphics[width=7.5cm]{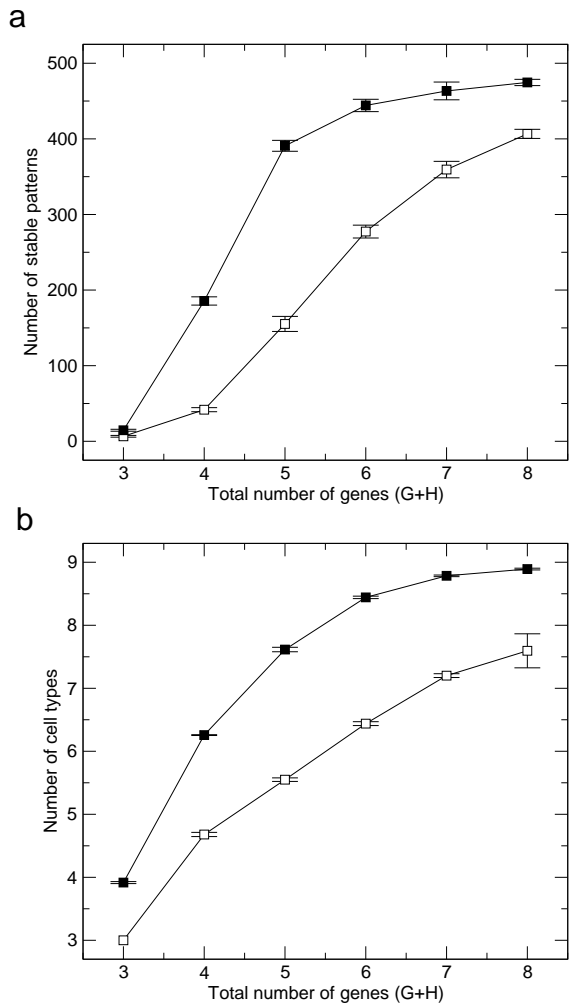}
\end{center}
\caption{
  \label{expl_fig}
  (a) Number of stable spatial patterns obtained for different
  combinations of genes and microhormones from the adaptive walk
  algorithm. Here two sets of numerical experiments were performed,
  involving one and two hormones and varying numbers of genes. In (b)
  the corresponding numbers of cell types are shown for the two cases. 
Note the saturation towards the maximum number $n_{ct}=C$ as the total 
number of genes increases. For each point, 5 simulations of 15000 generations were performed
  with $C=9$, $P=500$, and $I=100$.  }
\end{figure}

As an overall trend, simulations proceed in the direction of
increasing the number of patterns $|{\cal{P}}|$, as the average number
of cell types grows generation by generation. Given that every organism 
in the population performs its own adaptive walk, one would expect
the behavior of the average of the population would reflect some 
of the structure of the landscape. For instance, should there be any 
special local peaks in which an organism could get trapped, one would find 
many organisms there. As we will see, this is not the case.

A graph of $|{\cal{P}}|$
and the number of cell types achieved after 15000 generations for
different numbers of genes and hormones is shown in figure
\ref{expl_fig}. As the number of genes increases, the number of patterns
discovered increases, with a tendency to saturate at high $G$. Since the 
total number of possible patterns is, for $C=9$, $\vert \Gamma_g \vert =
2^9=512$, it is clear that the population almost finds all the possible
patterns for high $G$. This is even more evident if we look at the number 
of cell types, which reaches a value of almost 9, the maximum. From this curves 
it is also clear the simple change from one to two microhormones makes 
the system much less constrained and therefore more patterns are found.

\begin{figure}
\begin{center}
  \includegraphics[width=8cm]{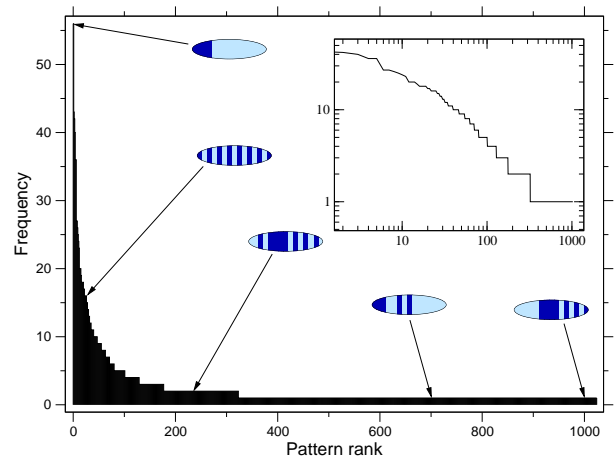}
\end{center}
\caption{
  \label{pattrankfig}
  Frequency-rank distribution of complex patterns obtained from the
  adaptive walk evolutionary algorithm, for 15000 generations, here
  with $G=3,H=2, P=500$ (a log-log plot is shown in the inset). A power decay is obtained, indicating a
  majority of patterns present in the final population but also a long
  tail of less common patterns.  Some examples are indicated. Most
  common patterns involve activation close to the anterior pole, but
  more complex patterns, such as stripes, appear to be rather
  frequent. As we move towards the tail of the distribution, more
  rich, asymmetric patterns are observed. }
\end{figure}

As mentioned earlier, given a population, many organisms generate some
patterns more than once. In fact, the distribution $F(r)$ of patterns
generated has an interesting structure, as shown in figure
\ref{pattrankfig}. Here patterns are sorted out by their rank $r$ 
(i. e. by their order from the most to the least frequent). Some patterns, especially
those with a small number of active cells at the anterior pole, are
very frequent. Some others, on the contrary, are very scarce. The
distribution follows approximately the form of a power law, i. e. $F(r) \sim r^{-\alpha}$, 
with $\alpha$ close to one. 

The frequent patters are, roughly, those that start the repeting
process that generates a wave across the organism but remain active
only near the anterior pole. Since less combinations exist for that
type of patterns, more organisms will generate the same ones.  The
rare patterns seem to be those that are the result of the construction
process and may vary considerably in its exact values, since many more 
different combinations exist. In figure \ref{temporal}, the pattern 
generated by $g_2$ would be an example of a ``starter'', and thus more 
frequent, and the $g_1$ would be an example of a possible rare one.

\begin{figure}
\begin{center}
  {\centering \includegraphics[width=8cm]{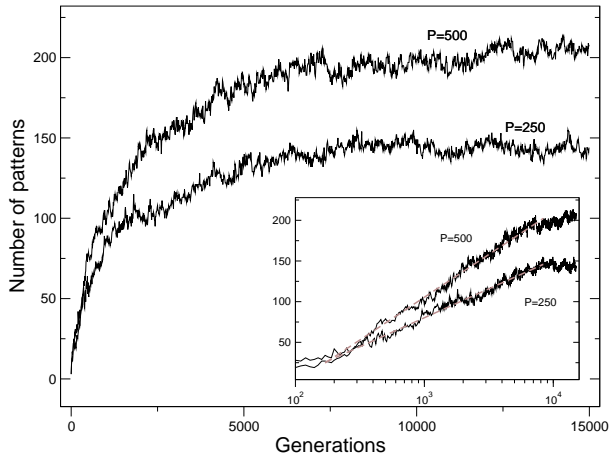} }
\end{center}
\caption{
  \label{comparedevolution}
  Time evolution of the number of stable patterns $\vert {\cal{P}}
  \vert$ for two different population sizes. The other parameters are
  $C=15$, $G=2$, $H=2$. The inset displays the same evolution in
  log-linear scale.  A straight line in such a plot indicates a
  logarithmic increase in the number of patterns, consistent with a
  rugged landscape.}
\end{figure}

\begin{figure}
\begin{center}
  {\centering \includegraphics[width=6cm]{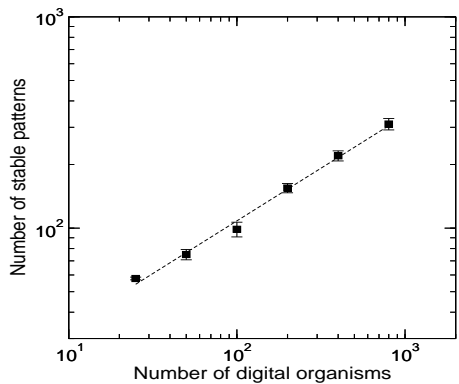} }
\end{center}
\caption{
  \label{numpatt_vs_P}
  The number of stable patterns $\vert {\cal{P}}
  \vert$ for different population sizes. The other parameters are
  $C=15$, $G=2$, $H=2$, and 15000 generations. The dashed line shows a 
  curve that fits the data with a power function, i.e.\ $f(x)=\alpha x^\beta$ 
  with an exponent of $\beta=0.498\pm0.02$.}
\end{figure}

Given a fixed $G$ and $H$, two other parameters affect the number of
patterns one obtains. On the one hand, the maximum number of
iterations to attain a stable pattern, $I$, and, on the other, the
number of organisms in the population, $P$. If $I$ is smaller, less
patterns are discovered since those patterns that may stabilize in a
higher number of iterations are filtered out. If $P$ is larger, the
landscape is explored more extensively, and more patterns are
discovered. Two figures demonstrate this dependence: figure \ref{comparedevolution} 
shows two different runs with different population sizes all other parameters being equal, 
showing how the bigger population attains a bigger number of patterns; figure 
\ref{numpatt_vs_P} shows the exact dependency. 

\begin{figure*}
\begin{center}
  {\centering 
    \includegraphics[width=11cm]{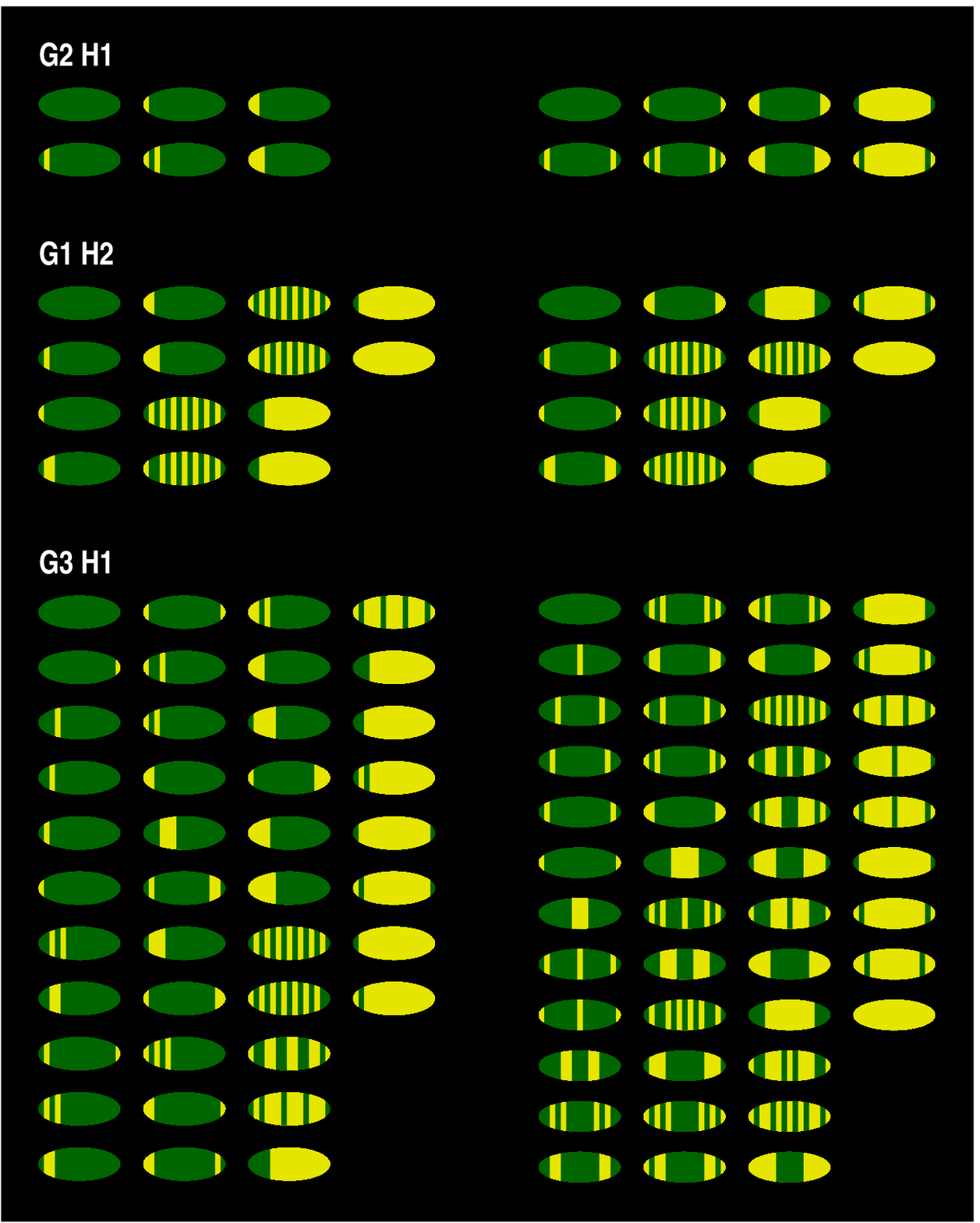} }
\end{center}
\caption{
  \label{patterns1}
  Taxonomy of patterns created from a system with different numbers of
  genes and microhormones at the subcritical regime. The left side
  shows the single pole case, whereas the right shows the symmetric
  one. A small amount of possible patterns is obtained, indicating
  that the possible repertoire of structures is fairly limited. Once
  two hormones are involved, it gets possible to obtain stable
  stripes, but only the most regular combinations are allowed.  The
  experiment involved a population of $P=500$ individuals for 15000
  generations.}
\end{figure*}

The most surprising results, however, involve the repertoire of spatial
patterns for small-sized gene networks. The basic set of patterns
obtained is shown in figure \ref{patterns1} for $C=15$ cells, where
the results from the two types of initial activations (one pole at
left, two poles at right) are shown.
As we can see, the repertoire is fairly limited, even if we use $G=3$
genes and $H=1$ hormones. Most patterns are rather homogeneous,
displaying a small gradient or single stripes, although some patterns
with wider stripes are also observable, thus indicating that the
richness of dynamical patterns transiently generated by coupled genes
can stably propagate even with only one hormone.

\begin{figure*}
\begin{center}
  \includegraphics[width=14.5cm]{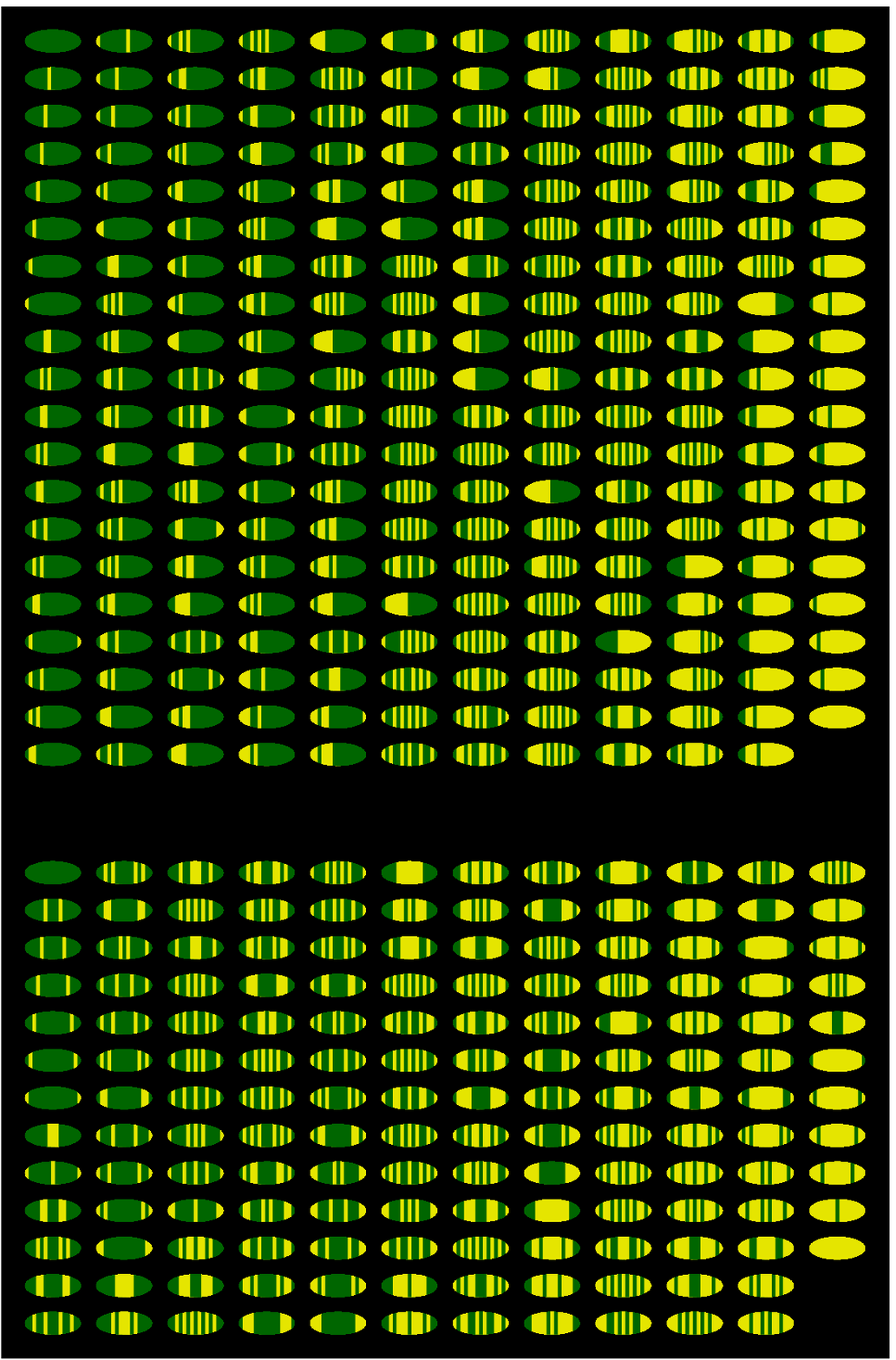}
\end{center}
\caption{
  \label{patterns2}
  As in figure \ref{patterns1}, but now using $G=2$ and $H=2$. 
  A much higher diversity of patterns is generated, revealing a combinatorial
  explosion of spatial motifs. The combination of complex patterns of
  gene interactions together with a combination of two microhormones
  leads to a great diversity of patterns of gene expression.  
  The upper part shows the patterns for a single-pole maternal signal, whereas 
  the lower part show the ones for the symmetric case.
}
\end{figure*}


The situation, however, dramatically changes once we cross a
complexity threshold. If $H=2, G=2$, a very large number of patterns
becomes available. In figure \ref{patterns2} the resulting stable
patterns are shown under the same simulation conditions. Now we can
see that the number of different patterns, $|{\cal{P}}_1|$=239 and
$|{\cal{P}}_2|$=154 for one and two maternal signals, respectively, is
comparable to the number of organisms involved. The diversity is also
remarkable: many different orderings in the stripe and gap
distributions are obtained, suggesting that a very high universe of
combinations is compatible with a high diversity of cell types.



\section{Discussion}

Any simple model of pattern formation does necessarily introduce shortcuts 
that limit the range of conclusions that can be safely reached. Our model 
does not include a large number of relevant features, from cell division to 
biologically realistic regulatory mechanisms. But again, as stressed in previous sections, 
some key, large scale features of complex systems do not depend on the specific details involved 
at lower levels \citep{Signs}. This is fairly well exemplified by Boolean models of 
{\em Drosophila} development. 

Our results can be summarized as follows (providing a tentative list of answers 
to the list of questions presented in the first section):

\begin{enumerate}

\item
Using the number of cell types as a complexity measure, our model indicates that 
the whole spectrum of spatial patterns is potentially reachable, provided that the 
population of digital organisms is large enough. Maximal diversity of cell types 
is positively correlated with the diversity of spatial patterns generated. This suggests 
that increasing cellular diversity is consistent with highly flexible 
pattern-forming mechanisms. 

\item
If a single hormone is involved, complex patterns can be obtained by increasing the 
network complexity. A much more rapid increase is obtained by using two hormones. Actually, 
the combination of two genes + two hormones leads to a combinatorial explosion of spatial 
patterns. As a consequence, our results indicate that, provided the number of organisms is 
large enough, any pattern seems to be available. 

\item
The spectrum of spatial patterns of gene expression is dominated by gap-like structures, 
stripes and all kind of combinations between them. It is remarkable that all classes 
of spatial structures are easily identified as matching the spatial distribution of maternal, 
gap and pair-rule gene expression patterns. Some genes (as it occurs in real development) are 
only transiently activated and thus appear in the end as absent. 

\item
The population climbs the underlying landscape in such a way that the rate 
of finding local peaks increases in a logarithmic fashion. This implies that a 
rapid diversification is followed by a further slowdown, in agreement with 
a rugged fitness landscape.

\item
The jump in pattern diversity experienced at $H=G=2$ indicates that thresholds 
in network complexity, even at small-gene numbers exist and can lead to combinatorial 
explosions. Such explosions would open a whole spectrum of available structures. Reaching 
such a threshold might require the formation of a minimal regulatory network and might also 
require other prerequisites dealing with body size, cellular interactions and tissue specialization. However, 
once in place, the whole universe of patterns can be made suddenly available.

\end{enumerate}

The previous results support the idea that the Cambrian event (and perhaps other rapid 
diversification events) might result from changes in the pattern of gene regulation. Of course 
the model does not consider ecological or other factors that might have played a 
key role. Instead, we concentrate in the study of the generative potential of such a simple 
model and in the universe of possible spatial patterns that can be generated. As a result, we 
obtain a surprising jump once a threshold of genetic complexity is reached. Although the 
whole repertoire of patterns might need some exploratory effort to be reached, most patterns 
are easily found and once a large fraction of them has been obtained, further innovation occurs 
at a slower pace. 

The main message of this paper is that rapid diversification events in terms of generation 
of evolutionary novelty in developmental processes can take place through combinatorial 
explosions. Although it can be argued that the model is too simple, the exploration of continuous 
counterparts of these results give very similar outcomes (particularly the thesholds of network 
complexity). Additionally, the continuous models allow to increase the repertoire of patterns 
and the same applies to discrete models using Boolean (instead of threshold) functions. In other words, the 
choices made here actually limit the diversity of patterns that can be generated.

\begin{acknowledgments}
  The authors would like to thank the members of the complex systems
  research group for useful discussions. This work was supported by
  a grant BFM2001-2154 (RVS) and the Generalitat de Catalunya (PFD, 2001FI/00732) and The
  Santa Fe Institute.
\end{acknowledgments}

\bibliographystyle{apsrmp}
\bibliography{evodevo}

\end{document}